\documentclass[]{spie}  

\usepackage{amsmath,amsfonts,amssymb}
\usepackage{graphicx}
\usepackage[colorlinks=true, allcolors=blue]{hyperref}

\title{Characterizing visual cortical magnification with topological smoothing and optimal transportation}

\author[a]{Yujian Xiong}
\author[a]{Yanshuai Tu}
\author[b,c,d,*]{Zhong-Lin Lu}
\author[a,*]{Yalin Wang}

\affil[a]{School of Computing and Augmented Intelligence, Arizona State University, Tempe, AZ, United States of America}
\affil[b]{Division of Arts and Sciences, New York University Shanghai, Shanghai, China}
\affil[c]{Center for Neural Science and Department of Psychology, New York University, New York, NY, United States of America}
\affil[d]{NYU-ECNU Institute of Brain and Cognitive Science, NYU Shanghai, Shanghai, China}

\authorinfo{Further author information: (Send correspondence to Yujian.Xiong)\\Yujian.Xiong: E-mail: yxiong42@asu.edu}

\begin{document} 
\maketitle

\begin{abstract}
Human vision has different concentration on visual fields. Cortical magnification factor (CMF) is a popular measurement on visual acuity and cortex concentration. In order to achieve thorough measurement of CMF across the whole visual field, we propose a method to measure planar CMF upon retinotopic maps generated by pRF decoding, with help of our proposed methods: optimal transportation and topological smoothing. The optimal transportation re-calculates vertex location in retinotopic mapping, and topological smoothing guarantees topological conditions in retinotopic maps, which allow us to calculate planar CMF with the proposed 1-ring patch method. The pipeline was applied to the HCP 7T dataset, giving new planar results on CMF measurement across all 181 subjects, which illustrate novel concentration behavior on visual fields and their individual difference.
\end{abstract}


\section{INTRODUCTION}
\label{sec:intro}  

Mammalian visual cortex contains multiple representations of different visual field area \cite{horton1991representation,cowey1964projection}. It has been an attracting topic to obtain retinotopic representation through activity and anatomy of visual cortex \cite{dougherty2003visual}. In many studies, Blood Oxygenation Level Development (BOLD) fMRI activation data was used to provide great analysis on retinotopic mapping \cite{dumoulin2008population, tootell1998functional}, along with the population receptive field (pRF) model (Fig. \ref{fig:1}), which estimates the center and size of the receptive field that is monitored by certain voxel on the cortical surface. Retinotopic maps generated by such method dramatically improved our knowledge of human visual system \cite{dumoulin2008population, benson2018bayesian, lerma2020validation, lage2020investigating}. However, since pRF model was designed to fit fMRI signal on the cortex, it is difficult to link the retinotopic maps with the actual visual behaviours of subjects, which inspire us to propose a new measurements to illustrate visual behaviour through retinotopic maps.

Human primary visual cortex is an ideal system to investigate the relationship of cortex anatomy and visual perception and retinotopic mapping gives the projection of 3D brain cortex onto its responsible area on visual field. Intuitively, a larger responsible area means less visual acuity for a fixed area of cortex, thus researchers had introduced cortical magnification factor(CMF), the ratio between the distances of two points on the cortical surface and the corresponding points in the visual field, as one of the measurements of this phenomenon \cite{cowey1974human, gauss1877theoria, lage2020investigating, bordier2015quantitative, qiu2006estimating}. CMF does not only shows the relationship between visual acuity and cortical structure, it also changes asymmetrically as a function of eccentricity and polar angle \cite{cowey1974human, harvey2011relationship, silva2018radial, kupers2021asymmetries}, and possess large individual difference \cite{ himmelberg2022linking, kupers2021asymmetries, song2022linking, benson2021cortical}, which made it a crucial task to obtain accurate and area-specific measurements of CMF.

Currently, template-based pRF methods are able to calculate the retinotopic maps to acquire visual cortex area and corresponding receptive center and receptive size from fMRI signals on a voxel-by-voxel basis. However, due to low signal-to-noise (SNR) ratio and low spatial resolution, typical pRF results are usually not topological \cite{dumoulin2008population, lage2020investigating, lerma2020validation}, especially in those area that is close to fovea. But a profound analysis on CMF and other visual-cortical measurements require the retinotopic maps to preserve local neighborhood geometric relationships, i.e. neighboring points on the cortical surface should have neighboring retinal visual coordinates \cite{wandell2007visual}, which called as topology preserving.

The direct results of such pRF methods with topological violations are that people often measure CMF approximately, as a function of single parameter (eccentricity or polar angle). And current studies are only able to show asymmetry by comparing summed large-scope CMF in different direction \cite{kupers2021asymmetries, silva2018radial}, lacking the illustration of detailed area-specific CMF behaviour in visual field, particularly near the fovea. On the other hand, in order to acquire such detailed measurements of CMF, it is necessary to fix topological violations in pRF results. As topological violations break the requirement of visual system's hierarchical organization, each visual area may not represents a certain portion of retina.

During the past few decades, many researchers had done significant efforts \cite{warnking2002fmri, bordier2015quantitative, zeidman2018bayesian, qiu2006estimating, lerma2021population, merkel2018spatial, merkel2020modulating, benson2018human} on improving the quality of fMRI recording, enhancing the accuracy and stability of pRF solution, correcting or reducing the topological violations. But the topology preserving was not fully guaranteed until our former work \cite{tu2021topological}, which fully solved the topological violations in the proposed topological-smoothing algorithm. This correction make it possible for us to do accurate visual-related quantification of retinotopic maps, such as CMF measurements. In order to measure CMF directly on the 2D planar visual field, here we propose the 1-ring patch method. 

Furthermore, we applied optimal transportation to utilize the projection from visual cortex to the 2D planar disk. By generalizing our prior work \cite{tu2020computing} to relocate the coordinate of each voxel on the 2D planar disk, we tested the performance of the conventional pRF solution \cite{benson2018human}, pRF with topological smoothing \cite{tu2021topological}, and pRF with optimal transportation. We managed to improve the accuracy and credibility when solve the visual coordinates of receptive center, by our proposed pipeline. The improved pRF solution gave us enough confidence and exactness on the final CMF measurements.

Currently, researchers had found that there exists an approximately negative correlation between pRF size and CMF in particular area on visual field \cite{harvey2011relationship, kupers2021asymmetries}. That is, locations that have better visual acuity, are usually accompanied with larger CMF and smaller pRF size. This phenomenon is intuitive but former findings were still in large-scale and could not tell the correlation in specific area on the 2D planar visual field. Through the proposed pipeline, we will be able to dig deep into the phenomenon by overlapping our planar CMF results onto the pRF solution, revealing the correlation which are not widely inspected before.

Our pipeline make it possible to measure CMF through BOLD based fMRI exactly on the 2D planar visual field, instead of approximation on some certain direction. Through this pipeline, we will be able to make illustration of CMF for each individual, which will help us to characterize people's visual concentration behaviour and locating their individual difference. For example, some subjects may have better visual acuity in the lower region under the fovea, while others may observe more clearly in the upper region above the fovea. For validation of our method, the pipeline was applied on real retinotopic data from the Human Connectome Project (HCP) \cite{benson2018human}, and achieved results in line with expectations. Our CMF measurement pipeline helps to address the problems about human visual concentration behaviours, and shed light on how to illustrate functional visual difference between healthy and unhealthy subjects.

\section{Materials and Methods}

\subsection{Human Retinotopic Data}
We use the Human Connectome Project (HCP) retinotopic data, which is publicly available at \\(\href{https://www.humanconnectome.org/study/hcp-young-adult}{https://www.humanconnectome.org/study/hcp-young-adult}) \cite{uugurbil2013pushing, van2013wu}. The pRF results and necessary stimuli and analysis package are available at (\href{https://osf.io/bw9ec/}{https://osf.io/bw9ec/}) \cite{benson2018human} or (\href{https://balsa.wustl.edu/study/show/9Zkk}{https://balsa.wustl.edu/study/show/9Zkk}) \cite{van2017brain}. This retinotopic dataset consists of retinotopy data from 181 subjects who participate in the retinotopic mapping experiments with 7T fMRI scanning. The pRF results of the published dataset contains fitted pRF results of retinotopy and atlases of cortical surface for each individual.

\subsection{Methods}

\begin{figure*}[ht]
\begin{center}
\begin{tabular}{c}
\includegraphics[width=.95\linewidth]{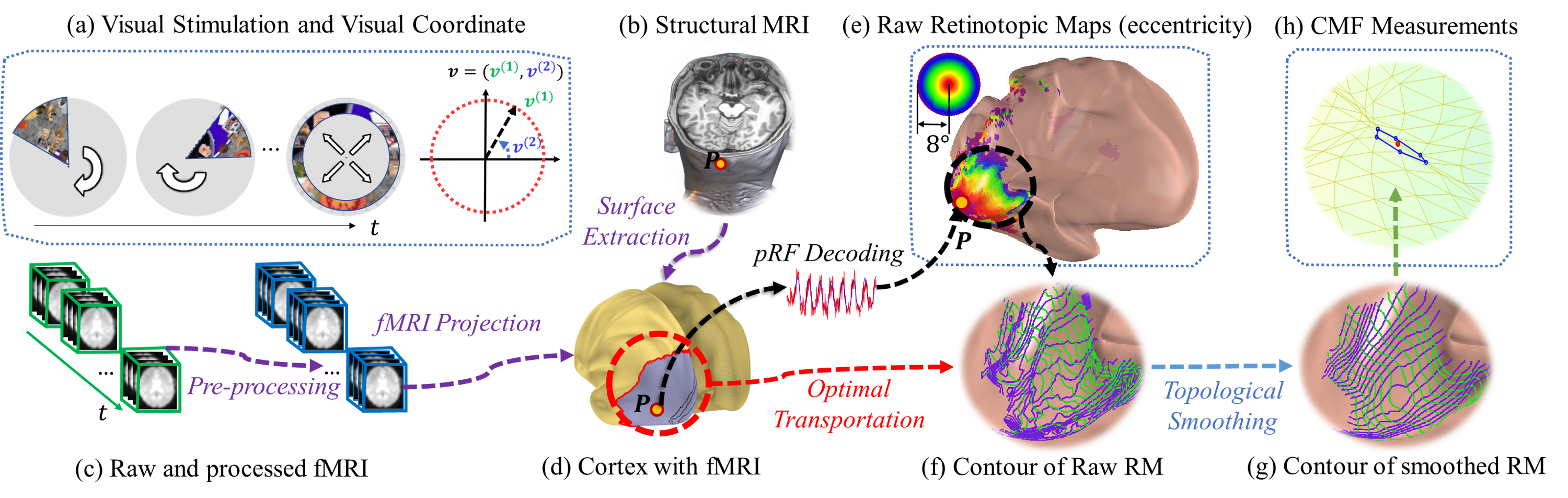}
\end{tabular}
\end{center}
\caption 
{ \label{fig:1}
\textbf{Overview of pipeline: } we start from two inputs, the first is (a) Visual stimuli and its coordinate system of the visual field, the second is (b) Structural (anatomical) MR image and (c) BOLD fMRI scans. These scanning sessions are pre-processed to fMRI signals, then we obtain (d) Reconstructed cortical surface with projected fMRI activation and surface extraction. After that, we calculate (e) retinotopic maps from the pRF model to acquire receptive center and pRF size $\sigma$ of each vertex. Further with projection of cortex onto a 2D planar disk with optimal transportation, we get clear retinotopic maps and (f) their contours. Finally, topological smoothing is applied to generalize the (g) smoothed retinotopic maps for (h) CMF measurements. } 
\end{figure*}

Besides receptive field size and population receptive field (pRF) size, cortical magnification factor (CMF) is also a popular measurement on visual acuity and cortex concentration. However, current methods calculate CMF by the ratio between the distance of two points on the cortical surface and the distance of corresponding points in the visual field. These linear measurements only exhibit trends of CMF under a narrow range of eccentricity but a wide range of polar angles. In order to achieve thorough measurement of CMF across the whole visual field, we propose a pipeline to obtain topological preserved retinotopic maps, and calculate planar CMF by dividing the area of certain patches on the cortical surface and corresponding area on the visual field with the proposed 1-ring patch method.  

The pipeline was applied to the V1, V2, and V3 area in Human Connectome Project 7T dataset. We analyzed 181 subjects from the HCP 7T MRI scans. Our pipeline contains fMRI pre-processing and projections onto cortex surface, then we apply compressive pRF decoding to obtain raw retinotopic maps. After that, we use optimal transportation to re-calculate the polar angle coordinates of each voxel on the flattened surface and use topological smoothing to fix topological violations in the retinotopic maps. The smoothed retinotopic maps are feasible for planar CMF measurement with 1-ring patch methods. Finally, we merge the results of the left and right hemisphere for each subject, in order to demonstrate the properties of CMF and their individual difference.

\subsection{Description of Purpose}
The mapping between visual field and cortical space indicates how each of us see the world. When a visual field is represented by a large area of cortical space, individual would have more concentration on this field, which could be interpreted as a larger acuity.

Typically, as eccentricity increases across visual field, neural receptive field (RF) and population receptive field (pRF) sizes increase, and cortical magnification factor (CMF) decreases, which imply a keen perception near the fovea and a coarser perception outside. There is also evidence on difference between perception performance across different polar angle. At certain eccentricity, we have observed different perception performance and asymmetries between left, right, upper, or lower vision area. Recent fMRI studies also suggest large difference in these perceptual behaviours between individuals.

To this extend, we propose a novel pipeline to clearly measure CMF on the whole visual field. The area-CMF in our work is defined as the area of a small patch on the cortex divided by the area of the patch correspondingly in visual field (with unit of $\bold{mm}^2 /\bold{deg}^2$), which is slightly different from the line-CMF (with unit of $\bold{mm} /\bold{deg}$) estimated based on the banded visual region.

We hypothesize that there may be variations in both pRF size and CMF across both eccentricity and polar angle, which could be interpreted as certain concentration area in the planar visual field. Our pipeline successfully measures the planar CMF, and shows different concentration area in each subjects' visual field.

\subsection{Pipeline Detail}
The proposed pipeline is shown in Fig.~\ref{fig:1}. In HCP dataset, the visual stimuli consist of rotating wedges, expanding or shrinking rings, and moving bars. The carriers of the stimuli (Fig.~\ref{fig:1}a) are made of dynamic color textures that can better activate corresponding neurons on visual cortex. We denote a point in the visual field by $\boldsymbol{v} = (v^{(1)}, v^{(2)}) \in \mathbb{R}^2$, where $v^{(1)}$ is the eccentricity (distance to the fovea in degrees of visual angle), and $v^{(2)}$ is the polar angle relative to the positive horizontal line (Fig.~\ref{fig:1}a). We also denote corresponding Cartesian coordinate of a point as $\boldsymbol{\rho} = (\rho_X, \rho_Y) \in \mathbb{R}^2$. Both high-quality structural MRI and fMRI scans were acquired by HCP group (Fig.~\ref{fig:1}b, Fig.~\ref{fig:1}c).

The pipeline starts with fMRI pre-processing, compressive population receptive field decoding. After that, we recalculated the parametric coordinate of each voxel by optimal mass transportation on the cortical surface, in order to modify the raw retinotopic maps (RM) from pRF decoding (Fig.~\ref{fig:1}e). Finally, there still exist many topological violations (Fig.~\ref{fig:1}f) if we draw a contour on the raw RM, so topological smoothing is applied to fix such topological violations, which make it possible for us to run 1-ring patch method for planar CMF measurement.

\begin{figure}[ht]
\begin{center}
\begin{tabular}{c}
\includegraphics[width=.6\linewidth]{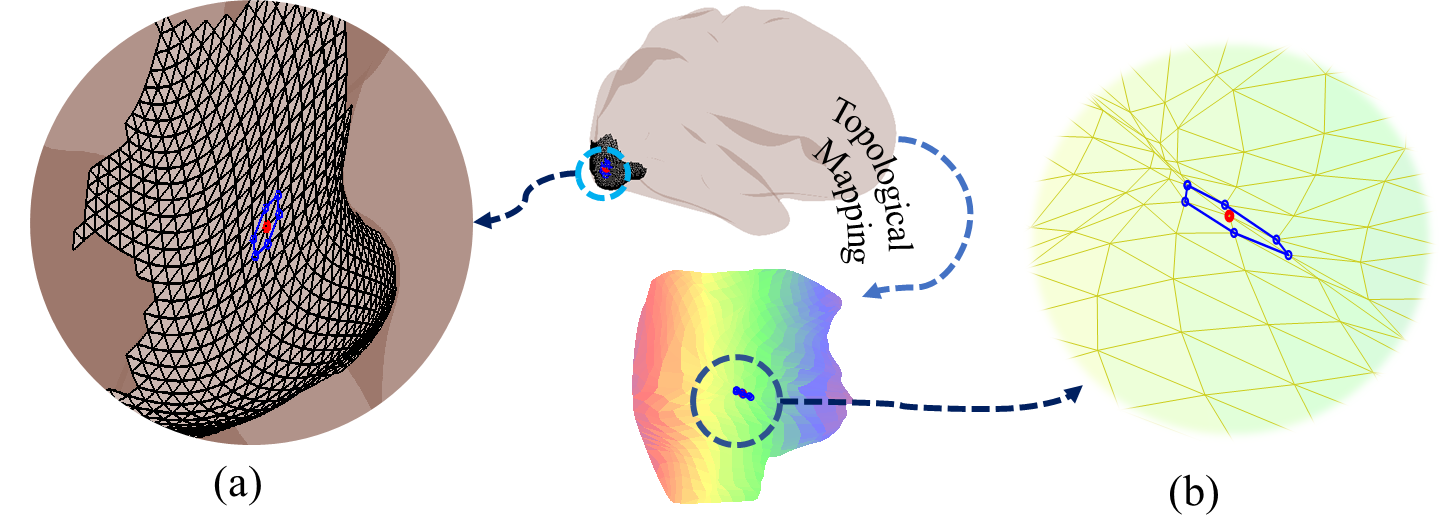}
\end{tabular}
\end{center}
\caption 
{ \label{fig:2}
1-ring patch measurement. (a) A vertex’s dual structure, (b) estimation of cortical magnification factor on the dual.} 
\end{figure}

The Cortical Magnification Factor (CMF) can be estimated based on a local structure, namely a vertex dual structure or the vertex 1-ring patch. The vertex 1-ring is a patch around the center vertex so that any vertex in the patch can reach the center vertex by “walking” within one edge. In Fig.~\ref{fig:2}a, we show the vertex (center red point) and its 1-ring patch (consisted of the enclosed blue polygon). After smoothing, each point on the cortical surface within the V1-V3 complex has a visual coordinate, i.e. we have a topological mapping for V1, V2, and V3. In Fig.~\ref{fig:2}b, we show that the 1-ring patch is also mapped as a 1-ring patch. Since topology is preserved, we can calculate the enclosed visual area. Eventually, the CMF for the center point is estimated by dividing the visual field area by cortical surface area within the vertex 1-ring patch. We shall mention that only a topological result can be used to estimate CMF in this way, otherwise, the visual area will be very noisy, that is the problem of non-topological results.

For outcome of our topological pipeline, we modeled the topological condition and generated topological and smooth retinotopic maps. The Beltrami coefficient, a metric of quasiconformal mapping, was used to define the topological condition. We developed a mathematical model to quantify topological smoothing as a constrained optimization problem and elaborated an efficient numerical method to solve it. As a result, we were able to improve boundary delineation. To our best knowledge, conventional methods have not fully considered topological constraints for multiple regions in retinotopic smoothing, which was not able for clear planar CMF measurements. While Our novel algorithm made the retinotopic maps from BOLD fMRI topological and consistent with results from neurophysiology, which improved the quality of retinotopic mapping and built a solid foundation for the 1-ring patch measurements.

\begin{table}[ht]
\centering
\caption{Comparison of CMF measurement methods}
\begin{tabular}{|l|l|l|}
\hline
Methods & \# of outlier point & Smoothing MSE \\
\hline
1. Linear CMF & 89 & 21.3 \\
\hline
2. Area Preserving & 267 & 18.7 \\
\hline
3. Proposed & 124 & 13.0 \\
\hline
\end{tabular}

\end{table}

In last step, we merged CMF result from both left and right hemisphere, to obtain a combined perceptual concentration on the full visual field.

\section{Results}

\subsection{CMF and Concentration Behaviour}
We have found that CMF of most subject centralized near fovea area, while their asymmetries exist across polar angle and differ from each other. We applied K-means clustering on the CMF results, obtained best elbow of K-value at 5. In these 5 clusters, people's concentration area lies separately in left-above, right-below, centralized, below-centralized, and some subjects shows 2 concentration area which we call dual-centralized.

\begin{figure}[ht]
\begin{center}
\begin{tabular}{c}
\includegraphics[width=0.65\linewidth]{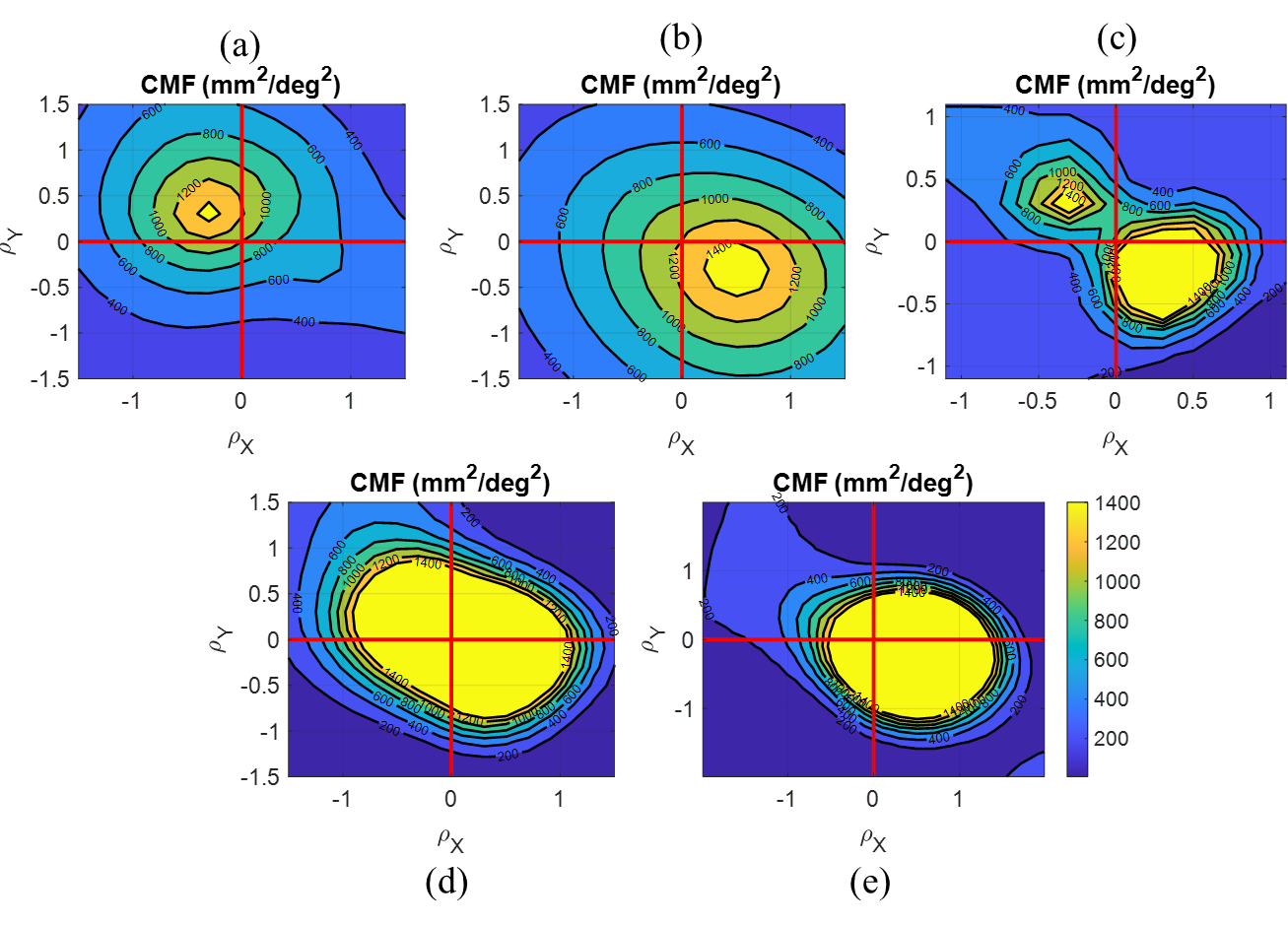}
\end{tabular}
\end{center}
\vspace*{-3mm}
\caption 
{ \label{fig:example2}
Example planar CMF result of 5 clusters: (a) left-above, (b) right-below, (c) dual-centralized, (d) centralized, (e) below-centralized. } 
\end{figure}

In former work, we had understood that most people have high focus around their fovea compared to outer visual field. However, in our work, we had shown that people preserve such focus but have different concentration behaviour near fovea.

\begin{table}[ht]
\centering
\caption{Number of subjects for each cluster}
\begin{tabular}{|l|l|l|}
\hline
Clusters & \# of subjects \\
\hline
a. left-above & 32 \\
\hline
b. right-below & 26 \\
\hline
c. dual-centralized & 11 \\
\hline
d. centralized & 47 \\
\hline
e. below-centralized & 68 \\
\hline
\end{tabular}

\end{table}

Most subjects (cluster d and e) have a large concentration area which decays steeply from the center to the outer ring. In this group, their behaviour are still different, where 68 people show an asymmetric concentration which has a offset downwards, and 47 people show an close-symmetric concentration that lies just at the fovea.

The other 69 subjects have a smaller concentration area which decays gently from the center to the outer ring. 32 of them locate their concentration towards the 2nd quadrant, 26 of them locate their concentration towards the 4th quadrant, while 11 of them show a strange behaviour which 2 concentration area were observed in opposite direction across the fovea.

To our knowledge, researchers haven't gone deep into the individual difference on the behaviour of concentration on visual field, the 5 different classes obtained by our CMF measurements show an interesting aspect on how people see the world, which would be quite difficult to understand through language. We look forward to future experiments which will go deep into this phenomenon.

\subsection{Correlation between pRF and CMF}
We also measured characters of pRF size across the visual field. Their performance corresponds with former work, and their asymmetries on polar angle align with our planar CMF results.

\begin{figure}[ht]
\begin{center}
\begin{tabular}{c}
\includegraphics[width=0.75\linewidth]{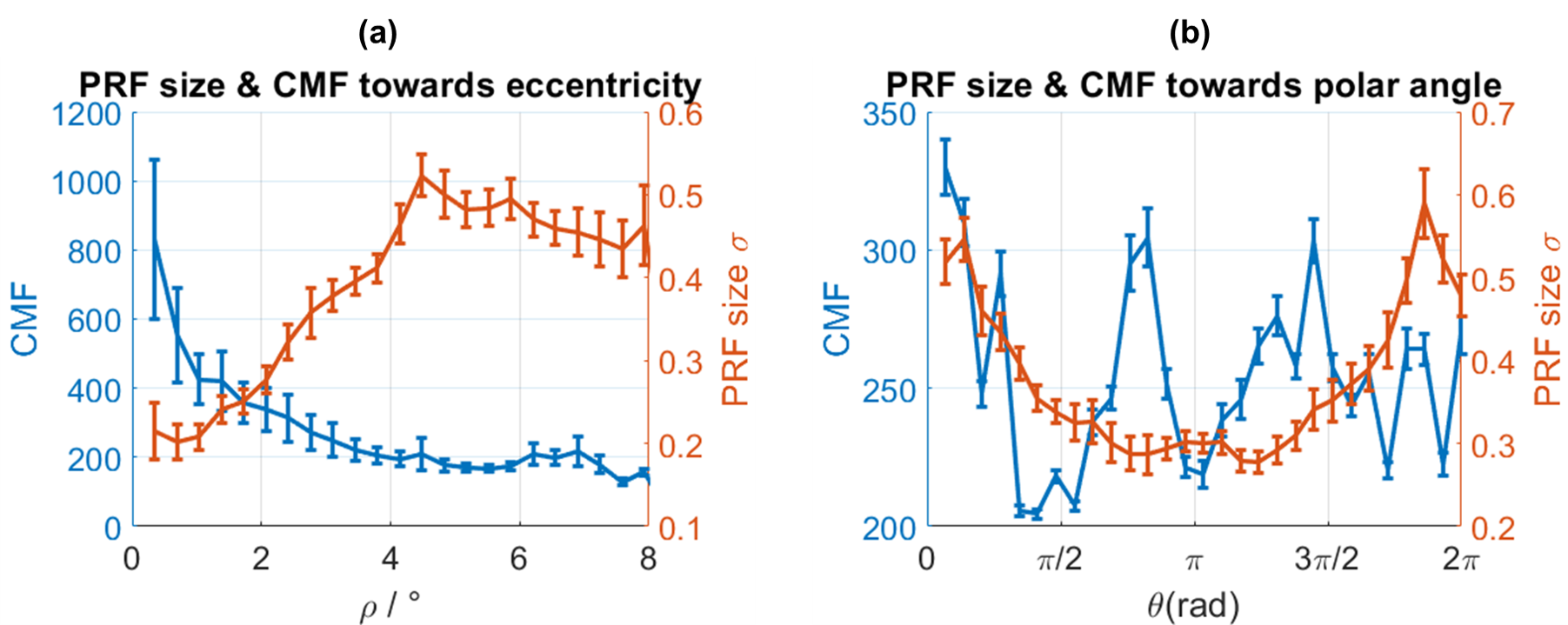}
\end{tabular}
\end{center}
\caption 
{ \label{fig:41}
 Correlation between CMF and pRF size across eccentricity $\rho$ and polar angle $\theta$ } 
\end{figure}

As we can see, there exists a negative correlation between pRF size $\sigma$ and CMF. As our point moves outwards from the fovea, its eccentricity would increase, which accompanies the increase of pRF size $\sigma$ and the decrease of CMF. This phenomenon accords with former experiment and intuition: our acuity in visual field reaches its highest at the fovea, and goes down for areas that is far away from the fovea.

As we can see, pRF size $\sigma$ and CMF spread evenly across different polar angle. This phenomenon accords with former experiment and intuition: our acuity in visual field reaches its highest at the fovea, and goes down for areas that is far away from the fovea.

To our knowledge, people had long supposed a negative correlation between pRF size and CMF across some variable such as eccentricity, polar angle, distance to fovea. However, with our methods, we could project the results of pRF decoding onto the visual field, where we could also calculate CMF on. To this extend, we will be able to analyze the correlation between pRF response and CMF in the 2D planar visual domain.

\subsection{Individual pRF and CMF correlation}
\begin{figure*}[ht]
\begin{center}
\begin{tabular}{c}
\includegraphics[width=0.95\linewidth]{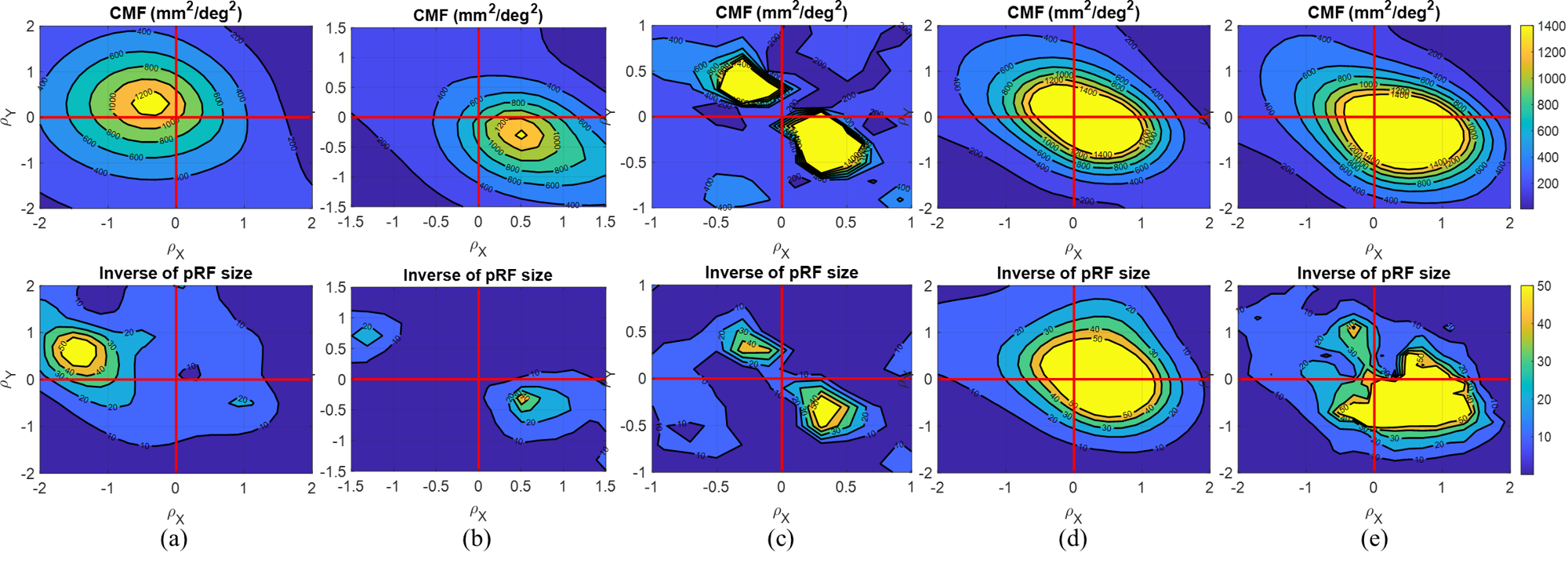}
\end{tabular}
\end{center}
\caption 
{ \label{fig:5}
Example comparison between planar CMF and inverse of pRF size $1/\sigma$: (a) Subject 4 (left-above), (b) Subject 10 (right-below), (c) Subject 170 (dual-centralized). (d) Subject 30 (centralized) (e) Subject 16 (below-centralized) }
\end{figure*}

Besides, we also analyzed population receptive field along with CMF in the subjects' visual field. The results was shown in Fig.~\ref{fig:5}. It showed that our CMF results give good alignment with the inverse of pRF size $1/\sigma$ around the 2D planar visual field.

As we known, smaller pRF size or larger CMF represents better acuity around particular point, and acuity decrease when eccentricity grew larger, which means points further from fovea would normally have higher pRF size and smaller CMF. In our work, we also illustrated the trend of pRF in the planar coordinate, which not only decrease along eccentricity growth, but also changes along polar angle.

We also notice that for most subjects, the inverse of pRF size $1/\sigma$ correlates with the CMF behaviour class. For example, We observe dual-centralized pattern in both planar CMF and $1/\sigma$ for subject 170. This phenomenon demonstrate that our clustering and assumptions could be tested in both CMF and pRF measurements, the individual behaviour difference could be seen in both 2 measurements.

\section{Discussion}
Compared to current methods, Our pipeline is able to measure planar CMF in 2D visual fields, could become a novel method to visualize human vision and concentration through the measurement of CMF. And suggests further experiment and analysis on various visual concentration across individuals.

We proposed a novel pipeline to measure planar CMF on the whole visual field from structure MRI and fMRI scan, our measurement and visualization shows that individuals have different concentration area. To our knowledge, the proposed topological smoother is the first method that guarantees topological conditions in retinotopic mapping, and based on this foundation, we made it possible for planar CMF measurement, which contributes a new gauging approach for future research on human vision and retinotopy.

Recently, many researches have shown asymmetries of cortical magnification in human visual cortex and their different performance across individuals with different twin status or gender. However, ordinary analysis showed the asymmetries but never got them clear enough. Our findings show that we could move one step further than measuring average CMF, but illustrating the planar CMF distribution to see exactly the asymmetries and their change across individuals.

Our results show that the individual difference of CMF is much more different than our previous thoughts. CMF of people varies in strength, concentration area, spreading. With our methods applied to different dataset in various scenes, we will be able to discover more variation of acuity and preference in human visual cortex.

\appendix    

\acknowledgments 
 
The first and second authors would thank Dr. Chengfeng Wen of Stony Brook University for providing fundamental geometry processing packages at the beginning of this project.

\bibliography{report} 
\bibliographystyle{spiebib} 

\end{document}